\pdfoutput=1 
\documentclass[epjCONF]{svjour}
\usepackage{amsmath} 
\usepackage{graphicx}
\usepackage[varg]{txfonts} 
\usepackage[latin1]{inputenc}
\session-title{Ultrafast Phenomena XVIII}
\newcommand{\tr}{\textrm} 
\newcommand{\mb}{\mathbf} 
\allowdisplaybreaks
\begin{document}
\title{Electron acceleration in vacuum by ultrashort and tightly focused radially polarized laser pulses}
\author{Vincent Marceau\thanks{\email{vincent.marceau.2@ulaval.ca}} \and Alexandre April \and Michel Pich\'e}
\institute{Centre d'optique, photonique et laser, Universit\'e Laval, Qu\'ebec, QC, G1V0A6, Canada}
\abstract{
Exact closed-form solutions to Maxwell's equations are used to investigate electron acceleration driven by radially polarized laser beams in the nonparaxial and ultrashort pulse regime. Besides allowing for higher energy gains, such beams could generate synchronized counterpropagating electron bunches.
} 
\maketitle
\section{Introduction\label{sec:TM01}}

The advent of ultra-intense laser facilities has led to exciting possibilities in the development of a new generation of compact laser-driven electron accelerators. Among the proposed laser acceleration schemes, the use of ultra-intense radially polarized laser beams in vacuum (termed \textit{direct acceleration}) is very promising, as it takes advantage of the strong longitudinal electric field at beam center to accelerate electrons along the optical axis~\cite{varin05_pre}. Numerical simulations have shown that collimated attosecond electron pulses could be produced by this  acceleration scheme~\cite{varin06_pre,karmakar07_lpb}.

Recent studies on direct acceleration have shown that reducing the pulse duration and beam waist size generally increases the maximum energy gain available~\cite{wong10_optexpress,wong11_optlett}. However, these analyses were carried under the paraxial and slowly varying envelope approximations. These approximations lose their validity as the beam waist size becomes comparable with the laser wavelength and the pulse duration approaches the single-cycle limit, conditions that are now often encountered in experiments . We propose a simple method to investigate direct acceleration in the nonparaxial and ultrashort pulse regime, and show that it offers the possibility of higher energy gains. We also highlight a peculiar feature of the acceleration dynamics under nonparaxial focusing conditions, namely the coexistence of forward and backward acceleration. This could offer a solution to the production of synchronized electron pulses required in some pump-probe experiments.

\section{Exact solution for a nonparaxial and ultrashort TM$_{01}$ pulsed beam\label{sec:TM01}}

Ultrashort and tightly focused pulsed beams must be modeled as exact solutions to Maxwell's equations. A simple and complete strategy to obtain exact closed-form solutions for the electromagnetic fields of such beams was recently presented by April~\cite{april10_intech}. For a TM$_{01}$ pulse, which corresponds to the lowest-order radially polarized laser beam, the field components are described by~\cite{april10_intech,marceau12_optlett}:
\begin{align}
&E_r (\mb{r},t) = \tr{Re}\, \bigg\{ \frac{3 E_0 \sin \tilde{2\theta}}{2\tilde{R}} \bigg( \frac{G_-^{(0)}}{\tilde{R}^2} - \frac{G_+^{(1)}}{c\tilde{R}} + \frac{G_-^{(2)}}{3c^2}\bigg) \bigg\} \ ,  \label{eq:npTM01Er}\\
&E_z (\mb{r},t) =  \tr{Re}\, \bigg\{ \frac{E_0}{\tilde{R}} \bigg[ \frac{(3\cos^2\tilde{\theta}-1)}{\tilde{R}}  \bigg( \frac{G_-^{(0)}}{\tilde{R}} - \frac{G_+^{(1)}}{c} \bigg)   - \frac{\sin^2\tilde{\theta}}{c^2} G_-^{(2)} \bigg] \bigg\}  \ , \label{eq:npTM01Ez} \\ 
&H_\phi (\mb{r},t) = \tr{Re}\, \bigg\{ \frac{E_0 \sin \tilde{\theta}}{\eta_0 \tilde{R}} \bigg( \frac{G_-^{(1)}}{c\tilde{R}} - \frac{G_+^{(2)}}{c^2}\bigg) \bigg\} \ . \label{eq:npTM01Hphi}
\end{align}
Here $E_0$ is an amplitude parameter, $\tilde{R}=[r^2 + (z+ja)^2]^{1/2}$, $\cos \tilde{\theta} = (z+ja)/\tilde{R} $, $G^{(n)}_\pm = \partial^n_t [f(\tilde{t}_+)\pm f(\tilde{t}_-)]$, $f(t) = e^{-j\phi_0}\left( 1- j \omega_0 t/s \right)^{-(s+1)}$, and $\tilde{t}_\pm = t \pm \tilde{R}/c + ja/c$. The function $f(t)$ is the inverse Fourier transform of the Poisson-like frequency spectrum of the pulse, in which $\omega_0 = c k_0$ is the frequency of maximum amplitude and $\phi_0$ is a constant phase~\cite{caron99_jmodoptic}. The parameter $a$, called the confocal parameter, is monotonically related to the beam waist size and characterizes the beam's degree of paraxiality: $k_0 a \sim 1$ for tight focusing conditions, while $k_0 a \gg 1$ for paraxial beams. The pulse duration $T$, which may be defined as twice the root-mean-square width of $|E_z|^2$, increases monotonically with $s$. In the limit $k_0 a \gg 1$ and $s \gg 1$, Eqs.~\eqref{eq:npTM01Er}--\eqref{eq:npTM01Hphi} reduce to the familiar paraxial TM$_{01}$ Gaussian pulse~\cite{fortin10_jpb}.

\begin{figure}[!t]
\centering
\includegraphics[width=0.32\textwidth]{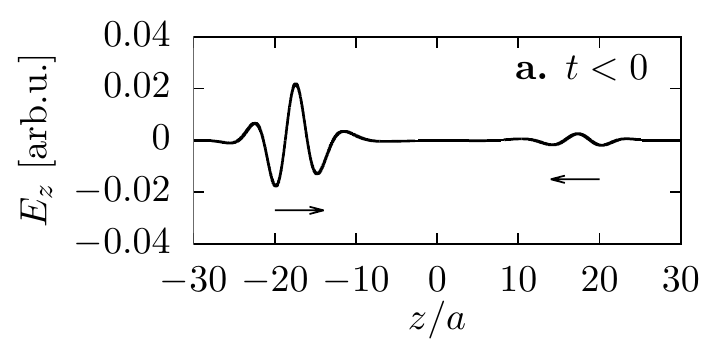} \
\includegraphics[width=0.32\textwidth]{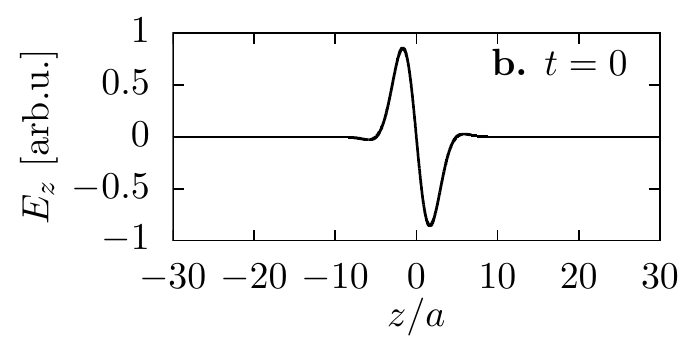} \
\includegraphics[width=0.32\textwidth]{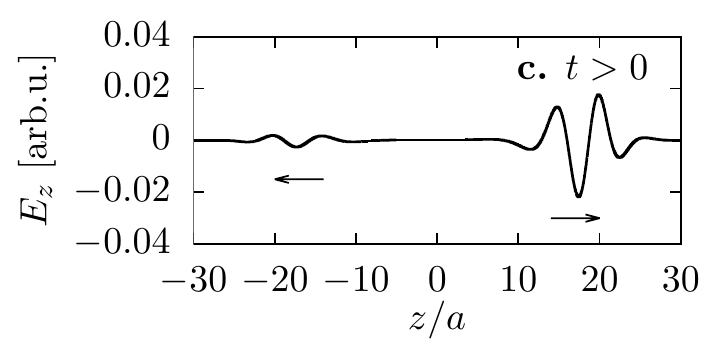} 
\caption{Longitudinal on-axis electric field of a TM$_{01}$ pulse with $k_0a = 1$ and $s=10$. \label{fig:1}}
\end{figure}

The TM$_{01}$ pulsed beam described above may be produced by focusing a collimated radially polarized input beam with a high aperture parabolic mirror. Its field distribution consists of two counterpropagating pulse components, as shown in Fig.~\ref{fig:1}~\cite{april10_optexpress}.

\section{On-axis acceleration in the nonparaxial and ultrashort pulse regime\label{sec:results}}

Direct acceleration is simulated by integrating the conventional Lorentz force equation for an electron initially at rest at position $z_0$ on the optical axis and outside the laser pulse. Since $E_r$ and $H_\phi$ vanish at $r=0$, the particle is accelerated by $E_z$ along the optical axis.

\begin{figure}[!b]
\centering
\includegraphics[width=0.45\textwidth]{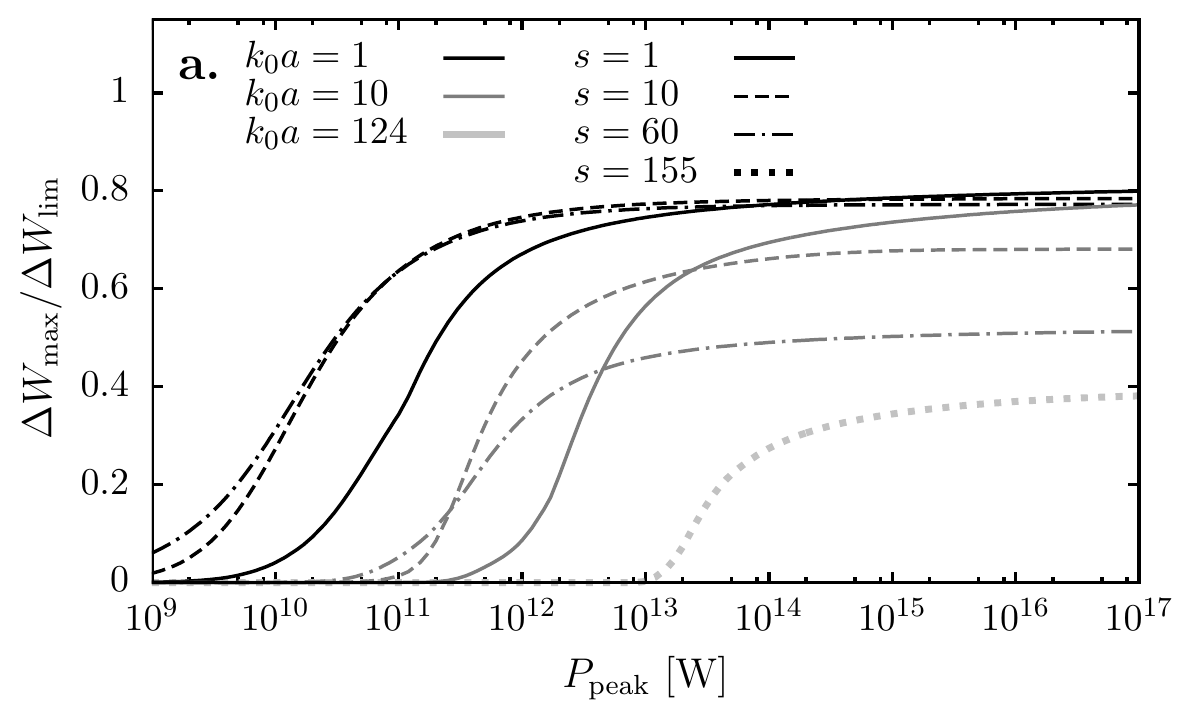} \ \ \
\includegraphics[width=0.45\textwidth]{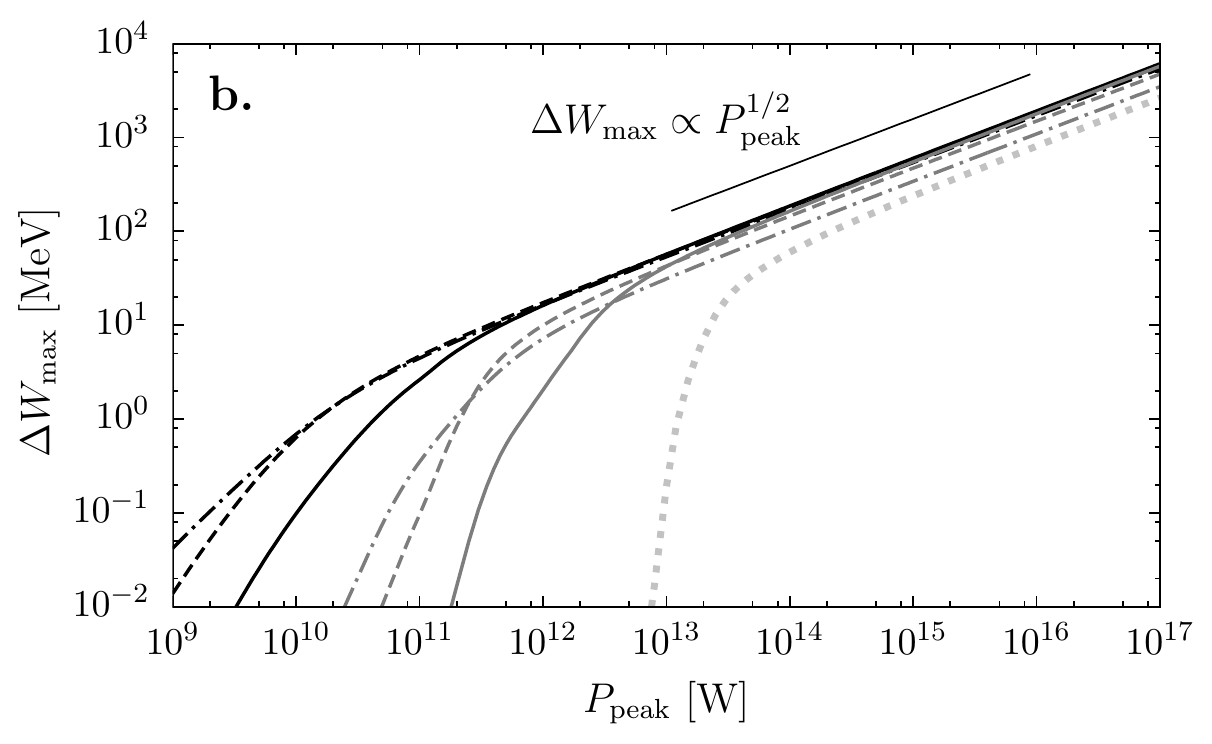} 
\caption{Maximum (a) normalized and (b) absolute energy gain of an electron initially at rest versus the laser pulse peak power for different values of $k_0 a$ and $s$. The curve with $\{k_0a=124,s=155\}$ corresponds to the limit of the paraxial regime investigated in~\cite{wong10_optexpress}. Figure taken from~\cite{marceau12_optlett}. \label{fig:2}}
\end{figure}

Figure \ref{fig:2} illustrates the variation of the maximum energy gain available $\Delta W_\tr{max}$ (after optimizing for $z_0$ and $\phi_0$) with the laser peak power $P_\tr{peak}$ for different combinations of $k_0a$ and $s$. Figure \ref{fig:2}a, in which $\Delta W_\tr{max}$ is expressed as a fraction of the theoretical energy gain limit $\Delta W_\tr{lim}$~\cite{fortin10_jpb}, shows that for constant values of $s$, the threshold power above which significant acceleration occurs is greatly reduced as $k_0 a$ decreases, i.e., as the focus is made tighter. According to Fig.~\ref{fig:2}b, MeV energy gains may be reached under tight focusing conditions with laser peak powers as low as 15 gigawatts. In constrast, a peak power about $10^3$ times greater is required to reach the same energy with paraxial pulses. At high peak power, Fig.~\ref{fig:2}a shows that shorter pulses yield a more efficient acceleration, with a ratio $\Delta W_\tr{max}/\Delta W_\tr{lim}$ reaching 80\% for single-cycle ($s=1$) pulses. Additional details about those results can be found in~\cite{marceau12_optlett}. 

\begin{figure}[!t]
\centering
\includegraphics[width=0.45\textwidth]{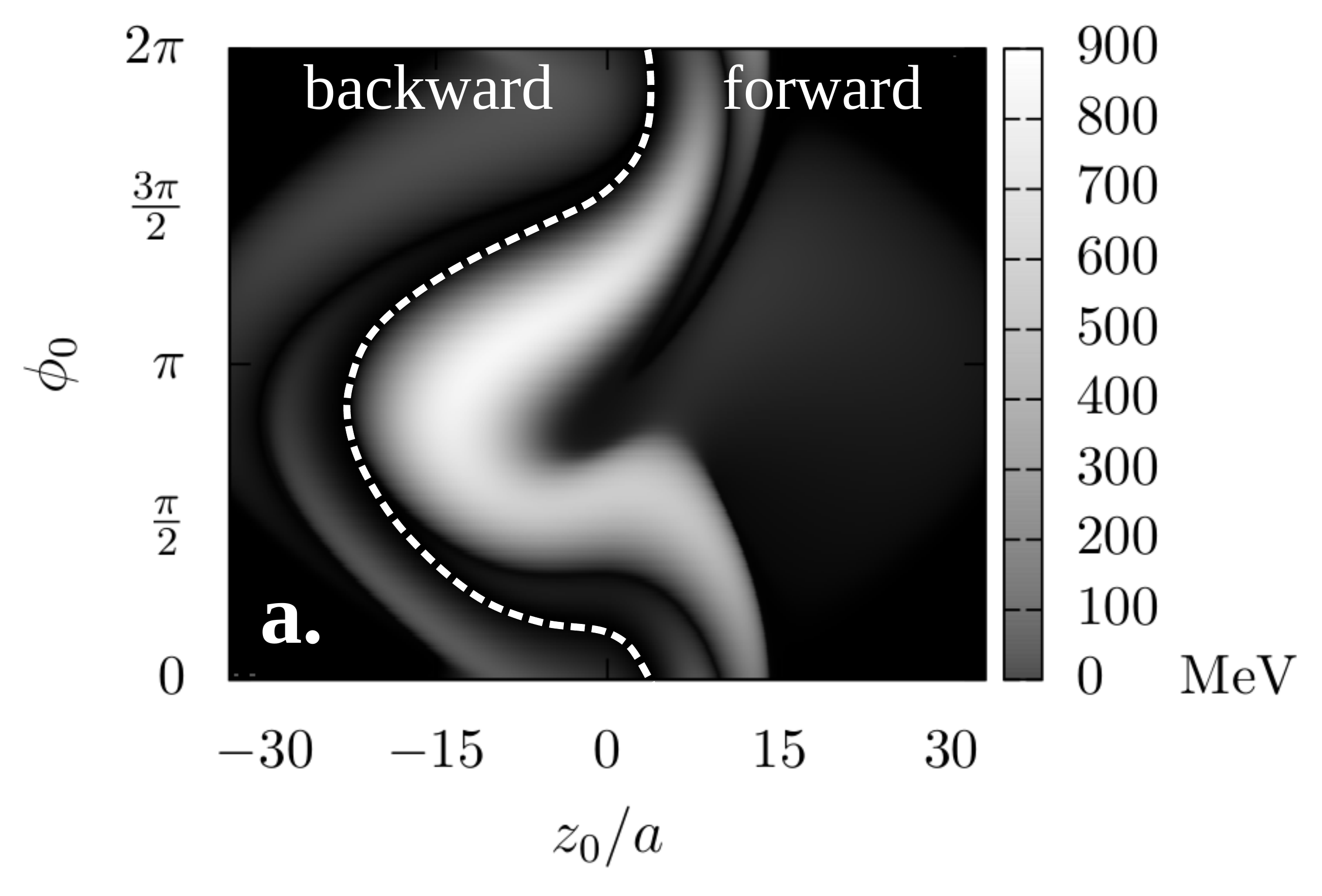} \ \ \
\includegraphics[width=0.35\textwidth]{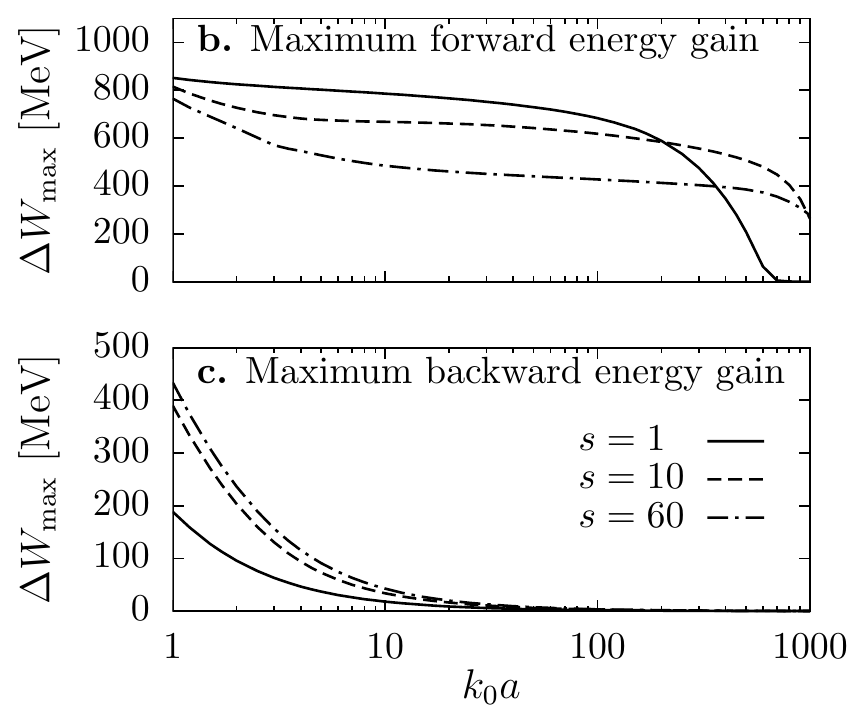}
\caption{(a) Energy gain of an electron initially at rest versus $z_0$ and $\phi_0$ for a laser pulse with $k_0 a = 1$, $s=1$. The dashed curve delimits the regions of forward and backward acceleration. (b)--(c) Maximum energy gain for electrons accelerated forward and backward versus $k_0 a$. In all figures, $P_\tr{peak}=2\times10^{15}$ W. \label{fig:3}}
\end{figure}

In the highly nonparaxial regime ($k_0 a \sim 1$), a closer look at the dynamics in the $(z_0,\phi_0)$ parameter space reveals the existence of two different types of acceleration (see Fig.~\ref{fig:3}a). In the first type, the electron is accelerated in the positive $z$ direction (forward acceleration), and may reach a high energy gain if its motion is synchronized with a negative half-cycle of the forward-propagating component of the beam. In the second type, the electron is accelerated in the negative $z$ direction (backward acceleration), and may similarly experience subcycle acceleration from the backward-propagating component of the beam. The maximum energy gain available from forward and backward acceleration is illustrated in Figs.~\ref{fig:3}b--c.  A significant backward acceleration is only observed under tight focusing conditions ($k_0 a < 10$), since the amplitude of the backward-propagating component of the laser beam rapidly decreases as $k_0 a$ increases.

\section{Conclusion \label{sec:conclu}}

We have highlighted the importance of going beyond the paraxial and slowly varying envelope approximations in the analysis of electron acceleration in vacuum by radially polarized laser beams. It was shown that the acceleration threshold power may be greatly reduced under tight focusing conditions, which demonstrates that direct acceleration is much more accessible to the current laser technology than previously expected. Moreover, our results hints that high-aperture focusing optics could be used to generate synchronized counterpropagating electron bunches. The proposed acceleration scheme could therefore find applications in the context of pump-probe experiments.

This research was supported by the Natural Sciences and Engineering Research Council of Canada, Le Fonds de Recherche du Qu{\'e}bec, and the Canadian Institute for Photonic Innovations.


\begin{thebibliography}{10}

\bibitem{varin05_pre}
C.~Varin, M.~Pich{\'e}, M.A. Porras, Phys. Rev. E \textbf{71}, 026603 (2005)

\bibitem{varin06_pre}
C.~Varin, M.~Pich{\'e}, Phys. Rev. E \textbf{74}, 045602(R) (2006)

\bibitem{karmakar07_lpb}
A.~Karmakar, A.~Pukhov, Laser Part. Beams \textbf{25}, 371 (2007)

\bibitem{wong10_optexpress}
L.J. Wong, F.X. K{\"a}rtner, Opt. Express \textbf{18}, 25035 (2010)

\bibitem{wong11_optlett}
L.J. Wong, F.X. K{\"a}rtner, Opt. Lett. \textbf{36}, 957 (2011)

\bibitem{april10_intech}
A.~April, in
  \emph{Coherence and Ultrashort Pulse Laser Emission}, F.J. Duarte ed.
  (InTech, 2010), pp. 355--382.

\bibitem{marceau12_optlett}
V.~Marceau, A.~April, M.~Pich{\'e}, Opt. Lett. \textbf{37}, 2442 (2012)

\bibitem{caron99_jmodoptic}
C.F.R. Caron, R.M. Potvliege, J. Mod. Opt. \textbf{46}, 1881 (1999)

\bibitem{fortin10_jpb}
P.L. Fortin, M.~Pich{\'e}, C.~Varin, J. Phys. B: At. Mol. Opt. Phys.
  \textbf{43}, 025401 (2010)

\bibitem{april10_optexpress}
A.~April, M.~Pich{\'e}, Opt. Express \textbf{18}, 22128 (2010)

\end{thebibliography}

\end{document}